\documentstyle[aps,prl,multicol]{revtex}
\begin{document}

\title{Analytic Coulomb matrix elements in the lowest Landau level in
disk geometry}

\author{E.V. Tsiper}

\address{Department of Chemistry, Princeton University, Princeton, NJ
08544$^*$}

\date{November 27, 2001}

\maketitle

 \begin{abstract}
 Using Darling's theorem on products of generalized hypergeometric
series an analytic expression is obtained for the Coulomb matrix
elements in the lowest Landau level in the representation of angular
momentum.  The result is important in the studies of Fractional
Quantum Hall effect (FQHE) in disk geometry.  Matrix elements are
expressed as simple finite sums of positive terms, eliminating the
need to approximate these quantities with slowly-convergent series.
As a by-product, an analytic representation for certain integals of
products of Laguerre polynomials is obtained.
 \end{abstract}

\section{Introduction}

The following integrals

\begin{eqnarray}
\label{Mmnl}
M_{mn}^l&=&(m+l,n|r_{12}^{-1}|m,n+l)\nonumber\\
&=&\int\int d^2r_1d^2r_2\psi^*_{m+l}({\bf r}_1)\psi^*_n({\bf r}_2)
 \frac{1}{|{\bf r}_1-{\bf r}_2|}
 \psi_m({\bf r}_1)\psi_{n+l}({\bf r}_2)
\end{eqnarray}

\begin{multicols}{2}

\noindent
 represent the Coulomb interaction matrix elements in the Lowest
Landau level.  These are the basic quantities for studies of
correlated two-dimensional systems in quantizing magnetic fields
\cite{ours_prl,ours_rho,girvin,jain,cappelli,hawrilak,kasner}.  The
single-particle wave functions in the angular momentum representation
are given by

\begin{equation}
\psi_m({\bf r})=(2\pi2^mm!)^{-1/2} r^me^{im\phi-r^2/4},
\label{psi}
\end{equation}
 where $r$ and $\phi$ are polar coordinates in the plane, and the
magnetic length $\sqrt{\hbar c/eH}$ is taken as a unit of length.  The
axial gauge for the vector potential ${\bf A}=\frac{1}{2}[{\bf H},{\bf
r}]$ is chosen.

Full Coulomb interaction has been shown to play crucial role in edge
effects in fractional quantum Hall systems, not captured by the
Laughlin's wave function\cite{ours_prl,ours_rho}.  The results of
Refs.~\onlinecite{ours_prl,ours_rho} would have been difficult to
obtain without an anlytic formulae for $M_{mn}^l$, derivation of which
is the subject of this work.  Use of well-known expressions by Girvin
and Jach\cite{girvin} is prohibitive at moderately large $m$ and $n$
because of large cancellations.  The problem was addressed
in\cite{stone}, where slowly-convergent series to approximate
$M_{mn}^l$ have been derived.

Here we present analytic formulae for $M_{mn}^l$ that contain simple
finite sums of positive terms, which can be easily evaluated for any
$m$, $n$, and $l$.  Moreover, the symmetry with respect to interchange
$m$ and $n$ is explicitly preserved.

\end{multicols}

\section{Formula for matrix elements}

We start with the result, which reads

\begin{equation}
M_{mn}^l=\sqrt{\frac{(m+l)!(n+l)!}{m!n!}}
  \frac{\Gamma(l+m+n+3/2)}{\pi2^{l+m+n+2}}
  \left[A_{mn}^lB_{nm}^l+B_{mn}^lA_{nm}^l\right],
\label{result}
\end{equation}
 where

\begin{mathletters}
\begin{eqnarray}
A_{mn}^l&=&\sum_{i=0}^m
\left(\begin{array}{c}
m \\
i
\end{array}\right)
\frac{\Gamma(1/2+i)\ \Gamma(1/2+l+i)}{(l+i)!\ \Gamma(3/2+l+n+i)},
\ \ \ \text{and}\\
B_{mn}^l&=&\sum_{i=0}^m
\left(\begin{array}{c}
m \\
i
\end{array}\right)
\frac{\Gamma(1/2+i)\ \Gamma(1/2+l+i)}{(l+i)!\ \Gamma(3/2+l+n+i)}
(1/2+l+2i).
\end{eqnarray}
\label{AB}
\end{mathletters}

\begin{multicols}{2}

The rest of the paper presents the derivation of Eqs.~(\ref{result})
and (\ref{AB}).  First, we substitute

\begin{equation}
 \frac{1}{|{\bf r}_1-{\bf r}_2|}=
 \int\frac{d^2q}{2\pi q}\exp[i{\bf q}({\bf r}_1-{\bf r}_2)]
\label{subs}
\end{equation}
 into Eq.~(\ref{Mmnl}).  The two separate integrals over ${\bf r}_1$
and ${\bf r}_2$ can be evaluated in terms of Laguerre polynomials
$L_m^l(q^2/2)$\cite{ryzhik}.  Substituting $q^2/2=x$ we obtain

\begin{equation}
M_{mn}^l=\sqrt{\frac{m!n!}{2(m+l)!(n+l)!}}
 \int\!dx\ x^{l-1/2}e^{-2x}L_m^l(x)L_n^l(x)
 \label{laguerre}
\end{equation}
 The above integral can be expressed\cite{prudnikov} using generalized
hypergeometric function \cite{bateman,koepf}

\begin{eqnarray}
M_{mn}^l&=&\sqrt{\frac{(l+m)!}{2\pi m!n!(l+n)!}}
 \frac{\Gamma(n+1/2)\ \Gamma(l+1/2)}{l!}\nonumber\\
 &\times&F\left(\left.
  \begin{tabular}{ccc}
    $1/2$, & $l+1/2$,  & $l+m+1$ \\
           & $-n+1/2$, & $l+1$
  \end{tabular}
  \right|
  \begin{tabular}{c}
    $z$\\
    \ 
  \end{tabular}
  \right)
\label{hypergeom}
\end{eqnarray}
 in the limit $z\rightarrow-1$, approached from the right.  The
function $F$ is defined as

\vskip 0.15 in

\hrule

\begin{equation}
 F\left(\left.
  \begin{tabular}{ccc}
    $a$, & $b$, & $c$ \\
         & $d$, & $e$
  \end{tabular}
  \right|
  \begin{tabular}{c}
    $z$\\
    \ 
  \end{tabular}
  \right)
 =\sum_{i=0}^\infty
   \frac{z^i(a)_i(b)_i(c)_i}{i!(d)_i(e)_i},
 \label{series}
\end{equation}
 where $(z)_i=z(z+1)...(z+i-1)=\Gamma(z+i)/\Gamma(z)$.  Taking the
limit avoids problems at $z=-1$, which is at the radius of convergence
of the power series (\ref{series}).  For $|z|<1$ the right-hand side
of Eq.~(\ref{hypergeom}) gives a more general integral of
$x^{l-1/2}e^{(z-1)x}L_m^l(-zx)L_n^l(x)$, analytic in $z$.

When one of the upper parameters is a negative integer the series
(\ref{series}) terminate yielding a finite sum.  At $z=-1$ we have

\begin{equation}
 F\left(\left.
  \begin{tabular}{ccc}
    $-k$, & $b$, & $c$ \\
          & $d$, & $e$
  \end{tabular}
  \right|
  \begin{tabular}{c}
    $-1$\\
    \ 
  \end{tabular}
  \right)
=\sum_{i=0}^k
\left(\begin{array}{c}
k \\
i
\end{array}\right)
\frac{(b)_i(c)_i}{(d)_i(e)_i}.
\end{equation}

Since none of the upper parameters of $F$ in Eq.~(\ref{hypergeom}) are
negative integers, the series is infinite.  However, it appears
possible to transform Eq.~(\ref{hypergeom}) in such a way that the
result contains only terminating hypergeometric series.  Using the
Darling's theorem on products\cite{bible}, the infinite series
Eq.~(\ref{hypergeom}) are brought into a sum of products of
generalized hypergeometric series that each have at least one negative
integer upper argument, therefore, representing a finite sum.  The
Darling's theorem for the function $F$ reads:

\end{multicols}

\begin{eqnarray}
 (1-z)^{a+b+c-d-e}
 F\left(\left.
  \begin{tabular}{ccc}
    $a$, & $b$, & $c$ \\
         & $d$, & $e$
  \end{tabular}
  \right|
  \begin{tabular}{c}
    $z$\\
    \ 
  \end{tabular}
  \right)
&=&\frac{e-1}{e-d}
 F\left(\left.
  \begin{tabular}{ccc}
    $d-a$, & $d-b$, & $d-c$ \\
           & $d$, & $d+1-e$
  \end{tabular}
  \right|
  \begin{tabular}{c}
    $z$\\
    \ 
  \end{tabular}
  \right)
 F\left(\left.
  \begin{tabular}{ccc}
    $e-a$, & $e-b$, & $e-c$ \\
           & $e-1$, & $e+1-d$
  \end{tabular}
  \right|
  \begin{tabular}{c}
    $z$\\
    \ 
  \end{tabular}
  \right)
\nonumber\\
&+&\frac{d-1}{d-e}
 F\left(\left.
  \begin{tabular}{ccc}
    $e-a$, & $e-b$, & $e-c$ \\
           & $e$, & $e+1-d$
  \end{tabular}
  \right|
  \begin{tabular}{c}
    $z$\\
    \ 
  \end{tabular}
  \right)
 F\left(\left.
  \begin{tabular}{ccc}
    $d-a$, & $d-b$, & $d-c$ \\
           & $d-1$, & $d+1-e$
  \end{tabular}
  \right|
  \begin{tabular}{c}
    $z$\\
    \ 
  \end{tabular}
  \right)
\label{theorem}
\end{eqnarray}

Using Eq.~(\ref{theorem}) and setting $z=-1$ in the end we obtain

\begin{eqnarray}
M_{mn}^l&=&\sqrt{\frac{(l+m)!}{\pi m!n!(l+n)!}}
 \frac{\Gamma(n+1/2)\ \Gamma(l+1/2)}
 {2^{l+m+n+1}(l+n+1/2)l!}\nonumber\\
&&\left\{
 (n+1/2)\ F\left(\left.
  \begin{tabular}{ccc}
    $-m$, & $1/2$,  & $l+1/2$ \\
           & $l+1$, & $l+n+3/2$
  \end{tabular}
  \right|
  \begin{tabular}{c}
    $-1$\\
    \ 
  \end{tabular}
  \right)
 F\left(\left.
  \begin{tabular}{ccc}
    $-n$, & $-n-l$,  & $-n-l-m-1/2$ \\
           & $-n-1/2$, & $-n-l+1/2$
  \end{tabular}
  \right|
  \begin{tabular}{c}
    $-1$\\
    \ 
  \end{tabular}
  \right)
\right.
\nonumber\\
&&\left.
 \ \ \ \ \ \ +\ l\ F\left(\left.
  \begin{tabular}{ccc}
    $-m$, & $1/2$,  & $l+1/2$ \\
           & $l$, & $l+n+3/2$
  \end{tabular}
  \right|
  \begin{tabular}{c}
    $-1$\\
    \ 
  \end{tabular}
  \right)
 F\left(\left.
  \begin{tabular}{ccc}
    $-n$, & $-n-l$,  & $-n-l-m-1/2$ \\
           & $-n+1/2$, & $-n-l+1/2$
  \end{tabular}
  \right|
  \begin{tabular}{c}
    $-1$\\
    \ 
  \end{tabular}
  \right)
\right\}
\label{darling}
\end{eqnarray}
 Further, we prove the following hypergeometric identity, valid for
any positive integer $k$:

\begin{equation}
 F\left(\left.
  \begin{tabular}{ccc}
    $-k$, & $-k-a$,  & $-k-b$ \\
          & $-k-c$, & $-k-d$
  \end{tabular}
  \right|
  \begin{tabular}{c}
    $-1$\\
    \ 
  \end{tabular}
  \right)
=\frac{(1+a)_k(1+b)_k}{(1+c)_k(1+d)_k}
 F\left(\left.
  \begin{tabular}{ccc}
    $-k$, & $1+c$,  & $1+d$ \\
          & $1+a$, & $1+b$
  \end{tabular}
  \right|
  \begin{tabular}{c}
    $-1$\\
    \ 
  \end{tabular}
  \right)
\label{rr}
\end{equation}

The proof is obtained by reversing the order of summation in the
(finite) sum, $i\rightarrow k-i$, using the symmetry of the binomial
coefficients with respect to this substitution, and noticing that
$(-k-a)_{k-i}=(-1)^{k-i}(1+a)_k/(1+a)_i$.  Using identity
Eq.~(\ref{rr}) to transform the second hypergeometric function in each
of the two terms in Eq.~(\ref{darling}), we get

\begin{eqnarray}
&&M_{mn}^l=\sqrt{\frac{(l+m)!(l+n)!}{m!n!}}
 \frac{\Gamma(l+m+n+3/2)}
 {\pi 2^{l+m+n+1}(l+n+1/2)}\nonumber\\
&&\left\{
\sum_{i=0}^m
\left(\begin{array}{c}
m \\
i
\end{array}\right)
\frac{\Gamma(1/2+i)\ \Gamma(1/2+l+i)}{(l+i-1)!\ \Gamma(3/2+l+n+i)}
\sum_{j=0}^n
\left(\begin{array}{c}
n \\
j
\end{array}\right)
\frac{\Gamma(1/2+j)\ \Gamma(1/2+l+j)}{(l+j)!\ \Gamma(3/2+l+m+j)}
\right.
\nonumber\\
&&\left.
+\sum_{i=0}^m
\left(\begin{array}{c}
m \\
i
\end{array}\right)
\frac{\Gamma(1/2+i)\ \Gamma(1/2+l+i)}{(l+i)!\ \Gamma(3/2+l+n+i)}
\sum_{j=0}^n
\left(\begin{array}{c}
n \\
j
\end{array}\right)
\frac{\Gamma(3/2+j)\ \Gamma(1/2+l+j)}{(l+j)!\ \Gamma(3/2+l+m+j)}
\right\}
\label{rrdarling}
\end{eqnarray}

The two terms in Eq.~(\ref{rrdarling}) can be brought together,
restoring the symmetry with respect to $m$ and $n$:

\begin{eqnarray}
M_{mn}^l&=&\sqrt{\frac{(l+m)!(l+n)!}{m!n!}}
 \frac{\Gamma(l+m+n+3/2)}
 {\pi 2^{l+m+n+1}(l+n+1/2)}
\sum_{i=0}^m
\sum_{j=0}^n
(l+i+j+1/2)\times
\nonumber\\
&&\left(\begin{array}{c}
m \\
i
\end{array}\right)
\frac{\Gamma(1/2+i)\ \Gamma(1/2+l+i)}{(l+i)!\ \Gamma(3/2+l+n+i)}
\left(\begin{array}{c}
n \\
j
\end{array}\right)
\frac{\Gamma(1/2+j)\ \Gamma(1/2+l+j)}{(l+j)!\ \Gamma(3/2+l+m+j)}.
\end{eqnarray}

\begin{multicols}{2}

Finally, regrouping the terms to split the double sum back into a
product of single sums while preserving the $m\leftrightarrow n$
symmetry we arrive at Eqs.~(\ref{result}) and (\ref{AB}).

We note that Eqs.~(\ref{result}) and (\ref{AB}) also represent a
useful analytic representation for the integrals of products of
Laguerre polynomials Eq.~(\ref{laguerre}).

\end{multicols}

\begin{references}

\bibitem[*]{} etsiper@princeton.edu;\ \ this work has been done at the
Department of Physics, SUNY at Stony Brook.

\bibitem{ours_prl} V.J. Goldman and E.V. Tsiper, Phys. Rev. Lett. {\bf
86}, 5841 (2001).

\bibitem{ours_rho} E.V. Tsiper and V.J. Goldman, Phys. Rev. B (2001),
in press.

\bibitem{girvin} S.M. Girvin and T. Jach, Phys. Rev. B {\bf 28} 4506
(1983).

\bibitem{jain} G. Dev and J.K. Jain {\bf 45}, 1223 (1992).

\bibitem{cappelli} A. Cappelli, C. Mendez, J. Simonin, and G.R. Zemba,
Phys. Rev. B {\bf 58}, 16291 (1998).

\bibitem{hawrilak} A. Wojs and P. Hawrilak, Phys. Rev. B {\bf 56},
13227 (1997).

\bibitem{kasner} M. Kasner and W. Appel, Ann. Phys. Leipzig {\bf 3}
433 (1994).

\bibitem{stone} M. Stone et al., Phys. Rev. B {\bf 45}, 14156 (1992).

\bibitem{ryzhik} I.M. Ryzhik and I.S. Gradshtein, {\em Table of
integrals, series, and products}, N.Y., Acad. Press (1965).

\bibitem{prudnikov} A.P. Prudnikov, Yu.A. Brychkov, and O.I. Marichev,
{\em Integrals and Series.  Special Functions} [in Russian], Moscow,
Nauka (1983).

\bibitem{bateman} H. Bateman and A. Erdelyi, {\em Higher
Transcendental Functions}, N.Y., McGraw-Hill (1953).

\bibitem{koepf} W. Koepf, {\em Hypergeometric Summation},
Braunschweig, Vieweg (1998).

\bibitem{bible} W.N. Bailey, {\em Generalized Hypergeometric Series},
in {\em Cambridge Tracts in Mathematics and Mathematical Physics},
No. 32, Ed. by. G.H. Hardy, and F.R.S.E. Cunningham, N.Y. and London
(1964).

\end{references}
\end{document}